\documentclass[%
 reprint,
superscriptaddress,
 amsmath,amssymb,
 aps,
 prl
]{revtex4-2}

\usepackage{graphicx}%
\usepackage{dcolumn}%
\usepackage{bm}%
\usepackage{color}
\usepackage{tensor}
\usepackage{mathtools}
\usepackage{siunitx}

\usepackage{derivative}
\usepackage{comment}

\usepackage{physics2}
\usephysicsmodule{ab}
\usephysicsmodule{op.legacy}
\usephysicsmodule{braket}
\usephysicsmodule{nabla.legacy}

\usepackage{here}
\usepackage{appendix}
\usepackage[colorlinks,linkcolor=blue,urlcolor=blue, citecolor=blue]{hyperref}

\newcommand{\dd}[1]{\odif{#1}\,} 

\usepackage{xcolor}

\usepackage{newtxtext}  
\usepackage{newtxmath} %

\usepackage{comment}

\DeclareMathOperator{\erfc}{erfc}

\begin{document}
\title{Single-shot conditional displacement gate between a trapped atom and traveling light}
\author{Seigo Kikura}
\email{seigokikura00@gmail.com}
\affiliation{Faculty of Science and Engineering, Waseda University, 3-4-1 Okubo, Shinjuku-ku, Tokyo 169-8555, Japan}
\author{Hayato Goto}
\affiliation{RIKEN Center for Quantum Computing (RQC), Wako, Saitama 351-0198, Japan}
\affiliation{Corporate Laboratory, Toshiba Corporation, Kawasaki, Kanagawa 212-8582, Japan}
\author{Fumiya Hanamura}
\affiliation{Centre for Quantum Technologies, National University of Singapore, 3 Science Drive 2, Singapore 117543}
\affiliation{Department of Materials Science and Engineering, National University of Singapore, Singapore}
\author{Takao Aoki}
\affiliation{Faculty of Science and Engineering, Waseda University, 3-4-1 Okubo, Shinjuku-ku, Tokyo 169-8555, Japan}
\affiliation{RIKEN Center for Quantum Computing (RQC), Wako, Saitama 351-0198, Japan}

\begin{abstract}
We propose a single-shot conditional displacement gate between a trapped atom as the control qubit and a traveling light pulse as the target oscillator, mediated by an optical cavity.
Classical driving of the atom synchronized with the light reflection off the cavity realizes the single-shot implementation of the crucial gate for the universal control of hybrid systems.
We further derive a concise gate model incorporating cavity loss and atomic decay, facilitating the evaluation and optimization of the gate performance.
This proposal establishes a key practical tool for coherently linking stationary atoms with itinerant light, a capability essential for realizing hybrid quantum information processing.
\end{abstract}
\maketitle

\emph{Introduction}---Traveling light offers a promising platform for quantum information processing, owing to its unique features such as high scalability and long-distance transmission capability~\cite{Bartolucci2023,AghaeeRad2025}.
Light is a natural host of continuous-variable (CV) encoding methods, enabling efficient quantum error-correcting codes---as exemplified by Gottesman-Kitaev-Preskill (GKP) code~\cite{Gottesman2001}---by harnessing the inherent redundancy of the Hilbert space.
However, the controllability of CV systems of light is generally more limited than that of discrete variable (DV) systems, e.g., qubits encoded in a photon, due to the inherently weak nonlinearity.

To overcome this limitation, hybrid systems consisting of oscillators and qubits have been explored, where qubits serve as control knobs of oscillators.
In particular, conditional displacement (CD)---qubit-controlled oscillator displacement---gates have been successfully demonstrated across various stationary oscillator platforms, including the motional states of trapped-ion systems~\cite{Haljan2005, Fluhmann2019, Campagne-ibarcq2020, Deneeve2022} and circuit quantum electrodynamics (QED)~\cite{van_Loock2008, Eickbusch2022}.
The CD gate enables universal control over the oscillator and is essential for both the generation and error correction of Gottesman-Kitaev-Preskill (GKP) codes, thereby serving as a key building block in hybrid quantum information processing. 

Unlike the above exemplary systems employ stationary oscillators, extending such hybrid control to traveling light fields remains highly challenging because light propagates incessantly at the speed of light.
One of the typical interactions between traveling light and a matter qubit is conditional rotation mediated by an optical cavity; the reflection off an optical cavity coupled to an atomic qubit induces a qubit-dependent phase shift to light~\cite{Duan2004,Wang2005}. 
Yet, the energy-conserving nature of this passive interaction prevents a universal control of CV-DV hybrid systems.
In principle, with the aid of displacement operations, implementable by using linear optics, a CD gate can be synthesized from conditional rotation gates~\cite{van_Loock2008, Dhara2025}, as demonstrated in circuit QED~\cite{Campagne-ibarcq2020, Eickbusch2022}.
In practice, multiple reflections of light are required for a single CD gate, leading to setup complications and accumulation of errors.

In this work, we propose a single-shot CD gate between a trapped atomic qubit and a traveling light pulse.
The interaction between them is mediated by an optical cavity, where the atom couples to the cavity mode that interfaces with the propagating field through a partially transmitting mirror.
By driving the atom simultaneously with the incidence of the target pulse, we can realize an atom-conditional displacement gate on the reflected light.
Our single-shot implementation eliminates the need for multiple optical operations, which would otherwise incur additional temporal overhead and optical loss; this makes it a fast and robust hybrid gate.
Our proposal fills a missing component for a hardware-efficient universal control, which can be achieved by single-qubit gates, beam-splitter operations, and CD gates~\cite{Eickbusch2022, Liu2024}, thereby accelerating the development of light-atom hybrid systems.

\begin{figure}[b]
    \centering
    \includegraphics[width=\linewidth]{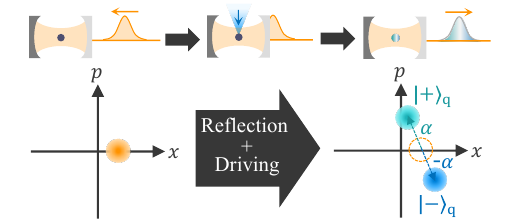}
    \caption{Schematic of the reflection-based conditional displacement (RCD) gate, along with the conceptual representation in the phase space.}
    \label{fig1}
\end{figure}

\emph{Reflection-based conditional displacement gate}---We first present the target operation that our proposed gate aims to realize, as shown in Fig.~\ref{fig1}.
The optical mode $\hat{a}$ represents a light pulse with a temporal envelope function $v(t)$, while qubit states $\{\ket{0}_\text{q}, \ket{1}_\text{q}\}$ are encoded in the internal states of an atom.
The atomic qubit is coupled to a one-sided cavity that interfaces with the external optical mode through a partially transmitting mirror.
To realize the desired atom-light gate, we incident the light pulse on the cavity with synchronized driving of the atom, resulting in a reflection-based conditional displacement (RCD) gate,
\begin{equation} \label{eq:ideal_RCD_gate}
    \text{CD}(\alpha) = \hat{D}(\alpha\hat{\sigma}_x) = e^{\hat{\sigma}_x (\alpha \hat{a}^\dagger - \alpha^\ast \hat{a})},
\end{equation}
where $\hat{\sigma}_x$ is the Pauli X operator on the qubit, and $\hat{D}(\alpha) = e^{\alpha \hat{a}^\dagger -\alpha\hat{a}}$ is the displacement operator.
In the phase space, this gate allows the qubit in $\ket{\pm}_\text{q} = (\ket{0}_\text{q} \pm \ket{1}_\text{q})/\sqrt{2}$ to displace the light state by $\pm\alpha$, as conceptually shown in Fig.~\ref{fig1}.

\begin{figure}[t]
    \centering
    \includegraphics[width=\linewidth]{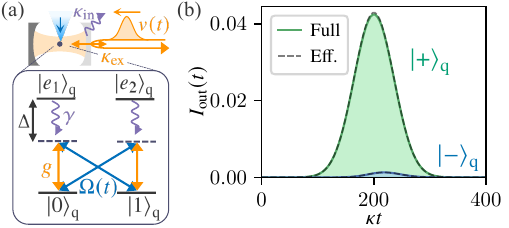}
    \caption{System and proof of concept for the RCD gate.
    (a) Detail of the cavity-QED system.
    The two optical transitions couple with a single cavity mode $\hat{c}$ at strength $g$, and the other diagonal transitions are driven by $\sigma_+$- and $\sigma_-$- polarized laser fields of Rabi frequency $\Omega(t)$ at the cavity frequency $\omega_c$.
    The detuning $\Delta = \omega_e-\omega_c$ of the atomic transition frequency $\omega_e$ is sufficiently large to suppress the atomic excitation.
    The one-sided optical cavity is coupled to the external optical field at rate $\kappa_\text{ex}$ while this also has a nonzero internal loss at rate $\kappa_\text{in} (< \kappa_\text{ex})$; the total cavity decay rate is $\kappa = \kappa_\text{ex} + \kappa_\text{in}$.
    The atom can decay spontaneously from excited states at rate $\gamma$.
    (b) State-dependent output-field intensity $I_\text{out}(t)$ via numerical simulation of the full model shown in panel (a) (solid lines) and the effective model (dashed lines).
    For the input coherent state $\ket{\beta=1}$ in a Gaussian pulse~\eqref{eq:gaussian_pulse}, centered at $t=4\tau$, applying the RCD gate $\text{CD}(\alpha = 1)$ ideally yields the coherent state $\ket{\beta = 2}$ ($\ket{\beta=0}$, the vacuum state) if the initial qubit state is in $\ket{+}_\text{q}$ ($\ket{-}_\text{q}$).
    The system parameters are $(\Delta, \kappa, \kappa_\text{in}, \gamma) = (20, 1, 0.01, 0.1)g$ and $\kappa \tau = 50$, resulting in $\epsilon_\text{pulse} = \num{8e-4}$ and the atomic decay probablity $p_\text{sp}\simeq 0.1$ [see Eqs.~\eqref{eq:epsilon_pulse}\eqref{eq:p_sp}].
    }
    \label{fig2}
\end{figure}

Hereafter, we present the details of the implementation in the cavity-QED system.
We consider a four-level system (FLS) with two stable and excited states, where the stable states act as the qubit states $\{\ket{0}_\text{q},\ket{1}_\text{q}\}$.
Here, we assume that the excited states can be adiabatically eliminated, by sufficiently large detunings and slow driving, leading to an effective Hamiltonian in the qubit space (its validity is discussed later).
In the rotating frame at the atomic transition frequency $\omega_e$ and an effective cavity frequency $\omega_c-\chi$, where $\chi = g^2/\Delta$ is the dispersive shift, the simplified Hamiltonian is given by [see Fig.~\ref{fig2}(a) for variable definitions and Sec.~\ref{sup_sec:system_model} for the derivation]
\begin{equation} \label{eq:H_sys^eff}
    \hat{H}^{\text{eff}}_{\text{sys}}(t) = \hat{\sigma}_{x}[\lambda(t) \hat{c}^{\dagger} + \lambda^\ast(t) \hat{c}],
\end{equation}
where $\hat{c}$ is the annihilation operator of the cavity field and
\begin{equation} \label{eq:lambda(t)_with_Ometa(t)}
    \lambda(t) = - \frac{g\Omega(t)}{\Delta}e^{-i\chi t}.
\end{equation}
Since the cavity mode couples to the output field mode at rate $\kappa_\text{ex}$, the total Hamiltonian is given by
\begin{equation}
    \begin{aligned}
        \begin{aligned}
            \hat{H}(t) =& \hat{H}_\text{sys}^\text{eff}(t) + \hat{H}_B + \hat{H}_\text{int}, \\
            \hat{H}_B =& \int \omega \hat{a}^\dagger(\omega) \hat{a}(\omega) \odif{\omega}, \\
            \hat{H}_\text{int} =& i\sqrt{\frac{\kappa_\text{ex}}{\pi}} \int  [\hat{a}^\dagger(\omega) \hat{c} - \hat{a}(\omega)\hat{c}^\dagger]\odif{\omega},
        \end{aligned}
    \end{aligned}
\end{equation}
where $\omega$ is the detuning of the field frequency from the effective cavity frequency $\omega_c-\chi$, and $\hat{a}(\omega)$ is a monochromatic annihilation operator in a propagating mode.
At a sufficiently large time $t=T$ to ensure the complete reflection, the unitary propagator of $\hat{H}(t)$ reduces to (see Sec.~\ref{ap_sec:RCD_gates} for the detailed derivation)
\begin{equation}
    \hat{U}(T) = \text{CD}(\alpha) e^{-i(\hat{H}_B + \hat{H}_\text{int})T},
\end{equation}
where we have set the time-dependent coupling as
\begin{equation} \label{eq:lambda(t)}
    \lambda(t) = \frac{i\alpha}{\sqrt{2\kappa_\text{ex}}} [\dot{v}(t)+\kappa v(t)].
\end{equation}
A byproduct unitary operator $e^{-i(\hat{H}_B + \hat{H}_\text{int})T}$, which represents an empty-cavity response, is translated into a $\pi$-phase shift for a sufficiently long pulse $v(t)$; this can be canceled by offsetting the phase origin by $\pi$ after the reflection.
Thus, the Rabi-type Hamiltonian~\eqref{eq:H_sys^eff} can realize the RCD gate.

We now incorporate the cavity internal loss at rate $\kappa_\text{in}$~\cite{Reiserer2015}, which can be modeled as the coupling of the cavity mode to a loss mode $\hat{a}_\text{loss}$ initially occupying the vacuum state $\ket{0}_\text{loss}$~\cite{Hacker2019, Hastrup2022}.
The system dynamics is obtained by tracing out the loss mode at the final step.
For a long pulse $v(t)$, the internal-loss effect is represented by an additional unitary operator 
\begin{equation}
    \hat{U}_\text{loss}(\alpha, \eta_\text{ex}) = \text{CD}_\text{loss}(\sqrt{\eta_\text{ex}^{-1}-1}\alpha)\hat{B}(\phi),
\end{equation}
with
\begin{equation}
    \cos\phi = 2\eta_\text{ex}-1, \quad \sin\phi = 2\sqrt{\eta_\text{ex}(1-\eta_\text{ex})},
\end{equation}
where $\eta_\text{ex}= \kappa_\text{ex}/\kappa$ is the coupling efficiency. 
The first beamsplitter operator $\hat{B}(\phi) = e^{\phi (\hat{a}\hat{a}_\text{loss}^\dagger - \hat{a}^\dagger \hat{a}_\text{loss})}$ between the desired and loss modes represents the optical loss due to the reflection.
The second conditional displacement operator $\text{CD}_\text{loss}(\sqrt{\eta_\text{ex}^{-1}-1}\alpha)$ represents the unintentional displacement of the loss mode.
The reflected state $\hat{\rho}(T)$, corresponding to the initial state $\hat{\rho}_\text{in}$ of the qubit and the desired-mode oscillator states, is given by sequentially applying the loss unitary operator and the ideal CD gate:
\begin{equation} \label{eq:no-atomic_decay_channel_1}
    \hat{\rho}(T) = \text{CD}(\alpha) \mathcal{E}(\hat{\rho}_\text{in}) \text{CD}^\dagger(\alpha),
\end{equation}
with 
\begin{equation} \label{eq:no-atomic_decay_channel_2}
    \mathcal{E}(\hat{\rho}) = \Tr_\text{loss}[\hat{U}_\text{loss}(\alpha, \eta_\text{ex}) (\hat{\rho} \otimes \ketbra{0}[_\text{loss}]{0}) \hat{U}_\text{loss}^\dagger(\alpha, \eta_\text{ex})].
\end{equation}

So far, we have considered a sufficiently long pulse $v(t)$; we now analyze a finite-length-pulse effect.
For simplicity, we consider a Gaussian pulse
\begin{equation} \label{eq:gaussian_pulse}
    v(t) = \frac{1}{(\pi \tau^2)^{1/4}}e^{-t^2/(2\tau^2)},
\end{equation}
where $\tau$ characterizes the pulse length.
First of all, the empty-cavity response, which is denoted by $\hat{B}(\phi)$ in a long pulse, indeed depends on the light frequency spectrum, disturbing the reflected pulse shape for large bandwidth (small $\tau$) light.
For a coherent state with average photon number $\bar{n}_\text{in}$, the deviation from $\hat{B}(\phi)$ can be calculated as (see Sec.~\ref{ap_sec:finite-pulse_effect})
\begin{equation} \label{eq:epsilon_pulse}
    \begin{aligned}
        \epsilon_{\text{pulse}} =& 1 - e^{-4\eta_\text{ex} \bar{n}_\text{in} [1-\sqrt{\pi} \kappa \tau e^{(\kappa \tau)^2} \erfc(\kappa\tau)]} \\
        =& \frac{2\eta_\text{ex} \bar{n}_\text{in}}{(\kappa \tau)^2} + \mathcal{O}\ab(\frac{1}{(\kappa \tau)^4}),
    \end{aligned}
\end{equation}
where $\erfc(x) = (2/\sqrt{\pi}) \int_x^{\infty}\dd{t} e^{-t^2}$ is the complementary error function.
Furthermore, a fast pulse induces a fast driving $\Omega(t)$, which will break down the adiabatic elimination assumed to derive the effective Hamiltonian~\eqref{eq:H_sys^eff}.
The elimination requires $\vab{\Delta} \gg g\sqrt{\aab*{\hat{c}^\dagger(t)\hat{c}(t)}}$ and $\vab{\Delta} \gg \vab{\Omega(t)}$.
The former condition is satisfied with a sufficiently large $\Delta$ for any $\tau$; for example, it reduces to $\vab{g\alpha/\Delta}^2 \ll \kappa \tau$ for $\bar{n}_\text{in}=0$~\cite{Kikura2025}.
The latter, however, does not depend on $\Delta$ because $\vab{\Omega(t)/\Delta} = \vab{\lambda(t)/g}$ from Eq.~\eqref{eq:lambda(t)_with_Ometa(t)}, yielding the requirement of the pulse width~\cite{Kikura2025}
\begin{equation}
    \frac{g^2}{\kappa}\tau \ab(1+ \frac{1}{\kappa \tau})^{-2} \gg \vab{\alpha}^2.
\end{equation}
A fast driving further increases the probability of the atomic spontaneous decay from excited states at rate $\gamma$.
The decay probability through an RCD gate is approximately given by (see Sec.~\ref{ap:atomic-decay_effect})
\begin{equation}
    p_\text{sp} \simeq 1- e^{\int \dd{t} 2\gamma_\text{eff}(t)},
\end{equation}
where the instantaneous effective decay rate is given by
\begin{equation}
    \gamma_\text{eff}(t) = \gamma \vab{\frac{\Omega(t)}{\Delta}}^2 = \gamma \vab{\frac{\lambda(t)}{g}}^2.
\end{equation}
From Eqs.~\eqref{eq:lambda(t)}\eqref{eq:gaussian_pulse}, we find
\begin{equation} \label{eq:p_sp}
    p_\text{sp} \simeq 1- \exp\ab\{-\frac{|\alpha|^2}{2\eta_\text{ex}(1-\eta_\text{ex})C_\text{in}}\ab[1+\frac{1}{2(\kappa\tau)^2}]\},
\end{equation}
where $C_\text{in} = g^2/(2\kappa_\text{in}\gamma)$
is the internal cooperativity.
Thus, to suppress the finite-pulse effect, the pulse length should satisfy the requirement
\begin{equation} \label{eq:pulse_length_requirement}
    \frac{g^2}{\kappa}\tau \gg |\alpha|^2, \kappa\tau \gg \max(1, 2\eta_\text{ex}\bar{n}_\text{in}),
\end{equation}
for a RCD gate with amplitude $\alpha$ on the input light with average photon number $\bar{n}_\text{in}$.

To confirm that the effective model---characterized by the effective Hamiltonian~\eqref{eq:H_sys^eff} and decay effect (see also Sec.~\ref{ap:numerical_simulation_method})---well captures the full-model dynamics under the pulse-length condition~\eqref{eq:pulse_length_requirement}, we simulate the RCD gate $\text{CD}(\alpha=1)$ on the coherent state $\ket{\beta=1}$, as an interesting example; the qubit $\ket{+}_\text{q}$ doubles the input amplitude, while the qubit $\ket{-}_\text{q}$ completely absorbs the input light, even though the system is driven by a classical laser.
The clear contrast of the output-laser intensity will enable an efficient qubit-state measurement via reflected-light power monitoring.
The qubit-dependent output-field intensity is presented in Fig.~\ref{fig2}(b), where we numerically simulate the full-level (solid lines) and effective (dashed lines) systems (see Sec.~\ref{ap:numerical_simulation_method} for detailed simulation methods); two results are in good agreement.

\emph{System optimization---}Our comprehensive analysis for gate performance, especially shown in Eqs.~\eqref{eq:no-atomic_decay_channel_1}\eqref{eq:p_sp}, clarifies two key metrics for characterizing system performance: the coupling efficiency $\eta_\text{ex}$ and the internal cooperativity $C_\text{in}$.
The former is readily tunable by changing the (effective) mirror transmittance; the tunability has been achieved in a fabrication process of an asymmetric diamond nanophotonic cavity~\cite{Knall2022}, and even dynamically in several cavity implementations, such as fiber-taper-coupled microsphere resonators~\cite{Volz2014}, nanofiber cavities~\cite{Kato2015}, and macroscopic resonators with one mirror placed outside a vacuum chamber~\cite{Shadmany2025}, to maximize the performance of their cavity-QED-based operations.
Thus, the latter metrics, $C_\text{in}$, should characterize intrinsic system performance, as in various protocols such as photon generation~\cite{Goto2019} and atom-photon gates~\cite{Goto2010, Borne2020, Kikura2025_caps}.

To efficiently optimize $\eta_\text{ex}$ for a fixed $C_\text{in}$, we note that a output state $\hat{\rho}_\text{fin}$ can be decomposed as $\hat{\rho}_\text{fin} \simeq (1-p_\text{sp}) \hat{\rho}_0 + p_\text{sp} \hat{\rho}_\text{sp}$, where $\hat{\rho}_0$ is given by the channel~\eqref{eq:no-atomic_decay_channel_1} and $\hat{\rho}_\text{sp}$ is the output state when the event with one or more atomic decays occur.
For an initial pure state $\ket{\psi_\text{ini}}$ that leads to the ideal final state $\text{CD}(\alpha)\ket{\psi_\text{ini}}$, the state fidelity $F$ satisfies
\begin{equation} \label{eq:inequality_of_F}
    F_\text{LB} \lesssim F \lesssim F_\text{LB} + p_\text{sp},
\end{equation}
where the approximated lower bound is given by
\begin{equation}
    F_\text{LB} = (1-p_\text{sp})\braket*[3]{\psi_\text{ini}}{\mathcal{E}(\ketbra{\psi_\text{ini}}{\psi_\text{ini}})}{\psi_\text{ini}}.
\end{equation}
Thus, with the analytical results for the channel $\mathcal{E}(\cdot)$ and $p_\text{sp}$ shown in Eqs.~\eqref{eq:no-atomic_decay_channel_2}\eqref{eq:p_sp}, we can easily optimize $\eta_\text{ex}$ by using an objective function such as $1-F_\text{LB}$, without resorting to exhaustive numerical simulations.

\begin{figure}
    \centering
    \includegraphics[width=\linewidth]{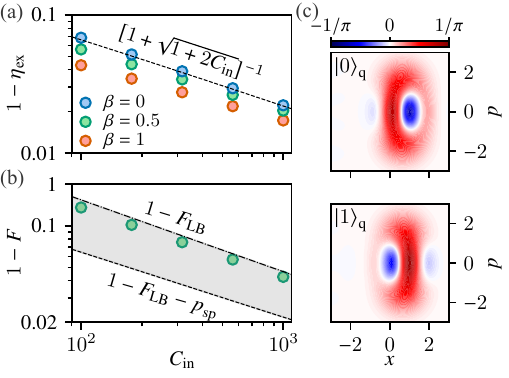}
    \caption{Optimization of the coupling efficiency $\eta_\text{ex}$ to maximize the gate performance.
    (a) Numerical optimization of $\eta_\text{ex}$ based on our analytical results~\eqref{eq:no-atomic_decay_channel_2}\eqref{eq:p_sp}, in the RCD gate $\text{CD}(1j)$ acting on $\ket{0}_\text{q}\ket{\beta}$.
    The dashed line represents $1-\eta_\text{ex} = 1/(1+\sqrt{1+2C_\text{in}})$, which captures well the result for $\beta=0$.
    (b) Infidelity $1-F$ for $\beta = 0.5$ with the optimized $\eta_\text{ex}$ [see panel (a)], where we numerically solve the master equation of an effective model, as explained in Sec.~\ref{ap:numerical_simulation_method}.
    Given $C_\text{in}$, the system parameters are $(\kappa_\text{in}, \Delta) = (0.01, 30)g$, which gives $\gamma = g^2/(2\kappa_\text{in} C_\text{in})$, while $\kappa_\text{ex}$ is determined by the optimized $\eta_\text{ex}$.
    The pulse length satisfies $g\tau = 300$.
    The dashed and dashdot lines respectively represent $1-F_\text{LB}-p_\text{sp}$ and $1-F_\text{LB}$ calculated by the analytical results~\eqref{eq:no-atomic_decay_channel_2}\eqref{eq:p_sp}.
    (c) Wigner functions of the output light with the postselection of the qubit states for $C_\text{in} = 1000$.
    Measuring the qubit state in $\ket{0(1)}_\text{q}$ projects the light state onto the non-classical state $\propto [\hat{D}(1j) \pm \hat{D}(-1j)]\ket{\beta=0.5}$ in the ideal case.
    Both of the projected states clearly show the Wigner negativity.
    }
    \label{fig3}
\end{figure}

As a demonstration, we optimize $\eta_\text{ex}$ in the RCD gate $\text{CD}(\alpha =1j)$ acting on $\ket{0}_\text{q}\ket{\beta}$ to minimize $1-F_\text{LB}$, as shown in Fig.~\ref{fig3}(a).
For $\beta = 0$, the optimized $\eta_\text{ex}$ is well approximated as $\eta_\text{ex} = 1 - 1/(1+\sqrt{1+2C_\text{in}})$, consistent with other protocols such as photon generation and atom-photon gates~\cite{Goto2019, Goto2010, Rosenblum2017, Kikura2025}.
As $\beta$ increases, the optimum value becomes larger to mitigate the optical loss induced by the internal cavity loss; even when $\alpha = 0$, the empty cavity acts as a loss channel with transmittance $\cos^2\phi = (2\eta_\text{ex}-1)^2$~\cite{Chuang1997}.
We further solve the master equation for $\hat{\rho}_\text{fin}$ and evaluate the infidelity, as shown in Fig.~\ref{fig3}(b) (see the detailed simulation method in Sec.~\ref{ap:numerical_simulation_method}).
This shows that the numerical solution $\hat{\rho}_\text{fin}$ satisfies inequality~\eqref{eq:inequality_of_F}, and that $F$ is close to $F_\text{LB}$, which indicates that an atomic-decay event results in nearly zero fidelity.
We also plot the Wigner functions with the postselection of the qubit states in $\ket{0(1)}_\text{q}$, shown in Fig.~\ref{fig3}(c).
The output states exhibit the Wigner negativity; this clearly shows that the RCD gate preserves qubit-light coherence through the gate dynamics.

Note that simulating the interaction between traveling and local quantum systems typically requires a numerical solution of the master equation in $\mathcal{N} \leq (N+1)\times d\times (M+1)$-dimension Hilbert space, where $N$ and $M$ are the maximum excitation numbers in the incoming and outgoing field, and $d$ is the dimension of the local system~\cite{Kiilerich2019}.
Thus, in particular, a large-photon-number state renders such numerical simulation time-consuming or even infeasible.
In the numerical simulations presented in Figs.~\ref{fig2}(b) and \ref{fig3}, we consider the initial light state to be a coherent state to make those feasible (see Sec.~\ref{ap:numerical_simulation_method}).
However, our analytical results~\eqref{eq:no-atomic_decay_channel_1}\eqref{eq:p_sp} satisfy for any initial state, thereby efficiently projecting gate performance, as shown in Fig.~\ref{fig3}(b).
Our analysis further indicates that achieving high gate performance primarily requires enhancing the internal cooperativity $C_\text{in}$.

\emph{Implementation---}We finally remark on potential physical implementations.
The idea of realizing the effective Hamiltonian~\eqref{eq:H_sys^eff} with the four-level system shown in Fig.~\ref{fig2}(a) was originally proposed for simulating the Dicke model~\cite{Dimer2007}, and was later demonstrated with $^{87}\text{Rb}$ atoms but acted as spin-1 states~\cite{Zhiqiang2017}.
Realizing our effective Hamiltonian with such alkali atoms should be possible~\cite{Takahashi2017}, while multiple Zeeman sublevels might cause additional errors.

More promising candidates that natively have well-separated four-level systems are diamond silicon-vacancy (SiV) centers and $^{171}\text{Yb}$ atoms.
SiV centers in a diamond nanophotonic cavity realized two ground and optically excited states with an appropriate magnetic field, enabling cavity-enhanced photon generation~\cite{Knall2022, Bersin2024}; this might straightforwardly realize our four-level system.
The other candidate, $^{171}\text{Yb}$, hosts a nuclear spin $1/2$ and offers a metastable qubit in $^{3}\text{P}_0, F=1/2$ states and optically excited states in $^3\text{D}_1, F=1/2$ states.
Cavity-QED systems with $^{3}\text{P}_0$-$^{3}\text{D}_1$ transition have recently been proposed along with concrete cavities such as silicon photonic crystal, twisted ring, and nanofiber cavities~\cite{Covey2019,Li2024, Sunami2025}, and a high-finesse nanofiber cavity was fabricated with projected cooperativity of 90~\cite{Horikawa2025}.
This four-level structure further gives the capability to mitigate the atomic decay effect by harnessing the versatile level structure of $^{171}\text{Yb}$; a fraction of the decay from $^{3}\text{D}_1$ states leaks outside the qubit subspace and eventually ends up in the ground state $^{1}\text{S}_0$, which can be detectable without disturbing the qubit coherence~\cite{Wu2022}.
This so-called erasure detection or conversion, which is applicable in RCD gates, enhances gate fidelity with postselection~\cite{Ma2023} and even significantly improves the threshold of fault-tolerant quantum computation~\cite{Wu2022,Sahay2023}, further enhancing the utility of RCD gates.

\emph{Conclusion}---We have proposed a single-shot implementation of an atom-conditional displacement gate on a traveling light pulse, where a four-level system (FLS) hosting a qubit is enclosed in an optical cavity.
Our proposal presents a hardware-efficient method; the single reflection of the itinerant pulse from the cavity, with the synchronized driving of the FLS, directly realizes a CD gate.
We have further analyzed the gate dynamics including the effects of cavity internal loss and atomic decay, and presented concise formulas to predict gate performance.
We have demonstrated that gate fidelity can be effectively captured by only two parameters: the coupling efficiency and the internal cooperativity.
This result streamlines the system optimization aimed at high performance, and thus enhances the utility of the single-shot CV-DV hybrid gate.
Optical CD gates are not only essential for quantum information processing but also have broad applications, such as the mitigation of optical loss~\cite{Park2022} and efficient entanglement distributions~\cite{Munro2008,Zhang2022,Macridin2025}; thus, our gate offers a practical route toward universal and scalable hybrid quantum computation and communication.
Furthermore, because this is implementable with a simple Rabi-type Hamiltonian, our proposal may be extended to other platforms such as circuit QED with traveling microwave pulses, offering a versatile approach for realizing hybrid quantum operations across diverse physical systems.

\section*{Acknowledgments}
This work was supported by JST Moonshot R\&D Grant Number JPMJMS2268.

\clearpage
\onecolumngrid

\begin{center}
\textbf{\large Supplemental Material for ``Single-shot conditional displacement gate between a trapped atom and traveling light''}

Seigo Kikura, Hayato Goto, Fumiya Hanamura, and Takao Aoki
\end{center}
\twocolumngrid

\setcounter{equation}{0}
\renewcommand{\theequation}{S\arabic{equation}}

\newcounter{supplsec}
\renewcommand{\thesupplsec}{S\arabic{supplsec}}

\newcommand{\supplsection}[1]{%
  \refstepcounter{supplsec}%
  \begin{center}
  \noindent\textbf{\thesupplsec: #1}\par
  \end{center}
}
\supplsection{System models} \label{sup_sec:system_model}

Here, we give full-level and effective models for a four-level system inside a cavity.
The full-level Hamiltonian under the rotating-wave approximation is given by~\cite{Dimer2007, Takahashi2017}
\begin{equation}
    \begin{aligned}
        &\omega_c \hat{c}^\dagger \hat{c} + \omega_e (\ketbra{e_1}[_\text{q}]{e_1} + \ketbra{e_2}[_\text{q}]{e_2}) \\
        &+ \Omega(t)(e^{-i\omega_c t}\ketbra{e_2}[_\text{q}]{0} + e^{-i\omega_c t}\ketbra{e_1}[_\text{q}]{1} ) + \text{h.c.} \\
        &+ g(\ketbra{e_1}[_\text{q}]{0} + \ketbra{e_2}[_\text{q}]{1} )\hat{c} + \text{h.c.},
    \end{aligned}
\end{equation}
where the laser frequency is set as the cavity one $\omega_c$.
In a rotating frame at frequency $\omega_c-\chi$, where $\chi$ is a free parameter, the Hamiltonian transforms as
\begin{equation} \label{ap_eq:full-level_H_sys}
    \begin{aligned}
        \hat{H}_\text{sys}(t) =& \chi \hat{c}^\dagger\hat{c} +  \hat{H}_e + \hat{V}(t) + \hat{V}^{\dagger}(t), \\
        \hat{H}_e =& (\Delta+\chi)  (\ketbra{e_1}[_\text{q}]{e_1} + \ketbra{e_2}[_\text{q}]{e_2}), \\
        \hat{V}(t) =& \Omega(t) e^{-i\chi t}(\ketbra{e_2}[_\text{q}]{0} + \ketbra{e_1}[_\text{q}]{1} ) \\
        &+ g(\ketbra{e_1}[_\text{q}]{0} + \ketbra{e_2}[_\text{q}]{1} )\hat{c},
    \end{aligned}
\end{equation}
with $\Delta = \omega_e-\omega_c$.
The adiabatic elimination of excited states gives an effective Hamiltonian up to the second order of $\hat{V}(t)$ as~\cite{Reiter2012, Kikura2025_purity}
\begin{equation}
    \begin{aligned}
        \hat{H}_\text{sys}^\text{eff}(t) =& \chi \hat{c}^\dagger\hat{c} - \hat{V}^\dagger(t) \frac{\ketbra{e_1}[_\text{q}]{e_1} + \ketbra{e_2}[_\text{q}]{e_2}}{\Delta + \chi} \hat{V}(t) \\
        =& \ab(\chi - \frac{g^2}{\Delta + \chi}) \hat{c}^\dagger\hat{c} - \hat{\sigma}_x \ab[\frac{g\Omega(t) e^{-i\chi t}}{\Delta + \chi}\hat{c}^\dagger + \text{h.c.}].
    \end{aligned}
\end{equation}
To cancel the first term, we set $\chi$ to satisfy $\chi = g^2/(\Delta+\chi)$, resulting in
\begin{equation} \label{ap_eq:H_sys^eff}
    \hat{H}_\text{sys}^\text{eff}(t) = \hat{\sigma}_{x}[\lambda(t) \hat{c}^{\dagger} + \lambda^\ast(t) \hat{c}],
\end{equation}
where 
\begin{equation}
    \chi = \frac{g^2}{\Delta}, \quad  \lambda(t) = - \frac{g\Omega(t)}{\Delta}e^{-i\chi t},
\end{equation}
by neglecting terms of third and higher order in $g/\Delta$.
Note that although we have assumed both the ground and excited states to be degenerate, our protocol should be robust against a small splitting, which is masked by a large detuning $\Delta$~\cite{Kikura2025}.

\supplsection{Channel of RCD gates} \label{ap_sec:RCD_gates}

We derive the unitary dynamics of the effective Hamiltonian.
The cavity mode $\hat{c}$ couples to the desired output mode $\hat{a}_0$ at rate $\kappa_\text{ex}$ and an unwanted mode $\hat{a}_1$ at rate $\kappa_\text{in}$, which consists of internal loss and scattering at mirrors.
Then, the total Hamiltonian is given by
\begin{equation}
    \begin{aligned} \label{ap_eq:total_eff_H}
        \hat{H}(t) =& \hat{H}_{\text{sys}}^\text{eff}(t) + \hat{H}_\text{B} + \hat{H}_\text{int}, \\
        \hat{H}_B =& \sum_{j=0,1} \int \omega \hat{a}_j^\dagger(\omega) \hat{a}_j(\omega) \dd {\omega}, \\
        \hat{H}_{\text{int}} =& \sum_{j=0,1} i\sqrt{\frac{\kappa_j}{\pi}} \int  [\hat{a}_j^\dagger(\omega) \hat{c} - \hat{a}_j(\omega)\hat{c}^\dagger] \dd{\omega},
    \end{aligned}
\end{equation}
where $\omega$ is the detuning of the field frequency from the effective cavity frequency $\omega_c-\chi$, and we have relabeled $ (\kappa_\text{ex}, \kappa_\text{in})$ as $(\kappa_0,\kappa_1)$ for notational simplicity and used the natural unit, $\hbar = c = 1$.
To derive the unitary propagator of $\hat{H}(t)$, we employ an appropriate unitary transformation with $\hat{U}_\text{d}(t)$ presented in Eq.~\eqref{ap_eq:U_d} or Ref.~\cite{Kikura2025}, giving a time-independent Hamiltonian 
\begin{equation} \label{ap_eq:H_d}
    \begin{aligned}
        \hat{H}_\text{d} =& \hat{U}_\text{d}^{-1}(t) \hat{H}(t) \hat{U}_\text{d}(t) + i\ab[\odv{\hat{U}_\text{d}^{-1}(t)}{t}] \hat{U}_\text{d}(t) \\
        =& \hat{H}_B + \hat{H}_\text{int},
    \end{aligned}
\end{equation}
which represents the empty cavity with resonant frequency $\omega_c-\chi$.
Thus, the unitary propagator is given by
\begin{equation}
    \hat{U}(t) = \hat{U}_\text{d}(t) e^{-i\hat{H}_\text{d} t}.
\end{equation}
In the following, we first analyze the empty-cavity dynamics $e^{-i\hat{H}_\text{d} t}$ and then present the detailed gate model.

\subsection{Reflection off an empty cavity} \label{subseq:reflection_empty_cavity}

We calculate the dynamics of $\hat{H}_\text{d}$ in the Heisenberg picture at $0\leq t \leq T$, where $T$ is an interaction-terminating time.
Using the input-output theory~\cite{Gardiner1985} gives the relation,
\begin{gather}
    \label{ap_eq:differential_c}
    \dot{\hat{c}} = \kappa \hat{c} - \sqrt{2\kappa_0}\hat{a}_{0,\text{out}}(t) -\sqrt{2\kappa_1}\hat{a}_{1,\text{out}}(t), \\
    \label{ap_eq:input_output_relation}
    \hat{a}_{0,\text{out}}(t) = \hat{a}_{0,\text{in}}(t) + \sqrt{2\kappa_0}\hat{c},
\end{gather}
where the field operators in the Heisenberg picture are given by
\begin{equation} \label{ap_eq:input/output_op_in_Heisenberg}
    \begin{aligned}
        \hat{a}_{j,\text{in}}(t) =& \frac{1}{\sqrt{2\pi}}\int\dd{\omega}e^{-i\omega t}\hat{a}_j(\omega, 0), \\
        \hat{a}_{j,\text{out}}(t) =& \frac{1}{\sqrt{2\pi}}\int\dd{\omega}e^{-i\omega (t-T)}\hat{a}_j(\omega,T), \\
    \end{aligned}
\end{equation}
with the operator $\hat{a}_j(\omega, \tau)$ at time $t = \tau$.
In the following, we may denote $\hat{a}_j(\omega, 0)$ as $\hat{a}_j(\omega)$.
Using the Fourier transform:
\begin{equation}
    \begin{aligned}
        f(t) =& \frac{1}{\sqrt{2\pi}}\int \dd{\omega} e^{-i\omega t}f(\omega), \\
        f(\omega) =& \frac{1}{\sqrt{2\pi}}\int \dd{t} e^{i\omega t}f(t),
    \end{aligned}
\end{equation} 
we obtain 
\begin{gather}
    \label{ap_ex:c(omega)}
    \hat{c}(\omega) = \frac{e^{i\omega T}[\sqrt{2\kappa_0}\hat{a}_0(\omega,T)+\sqrt{2\kappa_1}\hat{a}_1(\omega,T)]}{\kappa+i\omega}, \\
    \label{ap_ex:a_0(omega)}
    e^{i\omega T}\hat{a}_0(\omega,T) = \hat{a}_0(\omega) + \sqrt{2\kappa_0}\hat{c}(\omega),
\end{gather}
from Eqs.~\eqref{ap_eq:differential_c}\eqref{ap_eq:input_output_relation}, respectively.
Substituting Eq.~\eqref{ap_ex:c(omega)} into Eq.~\eqref{ap_ex:a_0(omega)} gives
\begin{equation} \label{ap_eq:input-output_empty_cavity}
    \hat{a}_0(\omega) = e^{i\omega T}[r^\ast(\omega)\hat{a}_0(\omega,T) +l^\ast(\omega)\hat{a}_1(\omega,T)],
\end{equation}
where the reflection coefficients are given by
\begin{equation} \label{ap_eq:r(omega)_l(omega)}
    r(\omega) =  \frac{\kappa -2\kappa_0 -i\omega}{\kappa-i\omega}, \quad l(\omega) =  - \frac{2\sqrt{\kappa_0\kappa_1}}{\kappa-i\omega},
\end{equation}
which satisfies $|r(\omega)|^2 + |l(\omega)|^2 = 1$.

From the above results in the Heisenberg picture, we will derive the reflected state for an arbitrary pure input state of a wavepacket mode $v$.
To describe such states in the Schr\"{o}dinger picture, we first define input and output field operators for the desired output mode in the Schr\"{o}dinger picture as
\begin{equation}
    \begin{aligned}
        \hat{a}_\text{in}^{\text{(s)}}(t) =& \frac{1}{\sqrt{2\pi}} \int \dd{\omega}e^{-i\omega t}\hat{a}_0(\omega), \\
        \hat{a}_\text{out}^{\text{(s)}}(t) =& \frac{1}{\sqrt{2\pi}} \int \dd{\omega}e^{-i\omega (t-T)}\hat{a}_0(\omega),
    \end{aligned}
\end{equation}
where we have used the superscript ``(s)'' to distinguish operators from those in the Heisenberg picture~\eqref{ap_eq:input/output_op_in_Heisenberg}.
We note that $\hat{a}_\text{in}^{\text{(s)}}(t) = \hat{a}_{0,\text{in}}(t)$ by definition.
We then define annihilation operators of a wavepacket mode $v$ as
\begin{equation}
    \hat{a}_\text{in(out)}[v] = \int \dd{t} v^\ast(t) \hat{a}_\text{in(out)}^{\text{(s)}}(t)
\end{equation}
which satisfies the bosonic commutation relations, such as $[\hat{a}_\text{in}[v], \hat{a}_\text{in}^\dagger[v]] = 1$, since $\int \dd{t} \vab{v(t)}^2 =1$.
These operators can be rewritten as
\begin{equation}
    \begin{aligned}
        \hat{a}_\text{in}[v] =& \int \dd{\omega} v^\ast(\omega) \hat{a}_0(\omega), \\
        \hat{a}_\text{out}[v] =& \int \dd{\omega} v^\ast(\omega) e^{i\omega T} \hat{a}_0(\omega).
    \end{aligned}
\end{equation}
An arbitrary pure input state of a wavepacket mode $v$ is given by $\ket{\Psi_\text{in}} = \sum_n c_n (\hat{a}_\text{in}^\dagger[v])^n \ket{\text{vac}}$, where $\ket{\text{vac}}$ is the vacuum state of all output modes.
By using the input-ouput relation~\eqref{ap_eq:input-output_empty_cavity}, we first rewrite 
\begin{equation}
    \begin{aligned}
    &\hat{a}_\text{in}^\dagger[v] \\
        &= \int \dd{\omega} v(\omega) \hat{a}_0^\dagger(\omega) \\
        &= \int \dd{\omega} v(\omega) e^{-i\omega T}[r(\omega)\hat{a}_0^\dagger(\omega,T) +l(\omega)\hat{a}_1^\dagger(\omega,T)], \\
        &= e^{i\hat{H}_\text{d} T}\int \dd{\omega} v(\omega) e^{-i\omega T}[r(\omega)\hat{a}_0^\dagger(\omega) +l(\omega)\hat{a}_1^\dagger(\omega)] e^{-i\hat{H}_\text{d}T} \\
        &= e^{i\hat{H}_\text{d}T} (\sqrt{\mathcal{N}_r}\hat{a}_\text{out}^\dagger[v_r] + \sqrt{\mathcal{N}_l}\hat{a}_\text{loss}^\dagger[v_l])e^{-i\hat{H}_\text{d}T},
    \end{aligned}
\end{equation}
where $v_r(\omega) = r(\omega)v(\omega)/\sqrt{\mathcal{N}_r}$ and $v_l(\omega) = l(\omega)v(\omega)/\sqrt{\mathcal{N}_l}$, with normalization constants $\mathcal{N}_r, \mathcal{N}_l$.
Here, we have defined the output field operator $\hat{a}_\text{loss}^{\text{(s)}}(t)$ in the same manner as $\hat{a}_\text{out}^{\text{(s)}}(t)$.
From $e^{-i\hat{H}_\text{d}T} \ket{\text{vac}} = \ket{\text{vac}}$, we find the final state in the Hamiltonian dynamics as
\footnote{The aurthors in Ref.~\cite{Hastrup2021} calculated the same dynamics with a similar method, but they mistakenly treated $\hat{a}_{0,\text{out}}(t)$ and $\hat{a}_{0,\text{out}}^{\text{(s)}}(t)$ as identical.} 
\begin{equation} \label{ap_eq:Psi(T)_empty_cavity}
    \begin{aligned}
        \ket{\Psi(T)} =& e^{-i\hat{H}_\text{d}T}\ket{\Psi_\text{in}} \\
        =& \sum_n c_n (\sqrt{\mathcal{N}_r}\hat{a}_\text{out}^\dagger[v_r] + \sqrt{\mathcal{N}_l}\hat{a}_\text{loss}^\dagger[v_l])^n \ket{\text{vac}}.
    \end{aligned}
\end{equation}

For a sufficiently long pulse $v(t)$, whose corresponding frequency spectrum is narrow compared to the cavity linewidth, the response function can be approximated by its on-resonance amplitude,
\begin{equation}
    r(\omega) \simeq r(0) = \frac{-\kappa_\text{ex} + \kappa_\text{in}}{\kappa}, \quad  l(\omega)\simeq l(0) = \frac{2\sqrt{\kappa_\text{ex}\kappa_\text{in}}}{\kappa},
\end{equation}
and thus the reflected state is given by
\begin{equation}
    \sum_n c_n \ab(\frac{-\kappa_\text{ex} + \kappa_\text{in}}{\kappa}\hat{a}_\text{out}^\dagger[v] + \frac{2\sqrt{\kappa_\text{ex}\kappa_\text{in}}}{\kappa}\hat{a}_\text{loss}^\dagger[v])^n \ket{\text{vac}}.
\end{equation}
The quantitative evaluation of this approximation is presented in Sec.~\ref{ap_sec:finite-pulse_effect}.
We further derive the unitary operator mapping the initial state to the final state.
To this end, we first define beam-splitter and phase-rotation operators of a wavepacket mode $v$ as follows:
\begin{equation}
    \begin{aligned}
        \hat{B}(\theta) =& e^{\theta(\hat{a}_\text{out}[v]\hat{a}_\text{loss}^\dagger[v]-\hat{a}_\text{out}^\dagger[v]\hat{a}_\text{loss}[v])}, \\
        \hat{R}_\text{out}(\theta) =& e^{i\theta \hat{a}_\text{out}^\dagger[v]\hat{a}_\text{out}[v]}.
    \end{aligned}
\end{equation}
By using the relations
\begin{equation}
    \begin{aligned}
        \hat{B}(\theta)\hat{a}_\text{out}[v]\hat{B}^\dagger(\theta) =& \hat{a}_\text{out}[v]\cos\theta + \hat{a}_\text{loss}[v]\sin\theta, \\
        \hat{R}_\text{out}(\theta)\hat{a}_\text{out}[v] \hat{R}_\text{out}^\dagger(\theta) =& \hat{a}_\text{out}[v]e^{-i\theta}, 
    \end{aligned}
\end{equation}
we find
\begin{equation}
    \begin{aligned}
        &\frac{-\kappa_\text{ex} + \kappa_\text{in}}{\kappa}\hat{a}_\text{out}^\dagger[v] + \frac{2\sqrt{\kappa_\text{ex}\kappa_\text{in}}}{\kappa}\hat{a}_\text{loss}^\dagger[v] \\
        &=  \hat{B}(\phi) \hat{R}_\text{out}(\pi) \hat{a}_\text{out}^\dagger[v] \hat{R}_\text{out}^\dagger(\pi) \hat{B}^\dagger(\phi)
    \end{aligned}
\end{equation}
with 
\begin{equation}
    \cos\phi = 2\eta_\text{ex}-1, \quad \sin\phi = 2\sqrt{\eta_\text{ex}(1-\eta_\text{ex})},
\end{equation}
where $\eta_\text{ex} = \kappa_\text{ex}/\kappa$ is the coupling efficiency.
The reflected state is rewritten as
\begin{equation}
    \begin{aligned}
        \ket{\Psi(T)} =  \hat{B}(\phi) \hat{R}_\text{out}(\pi) \sum_n c_n  (\hat{a}_\text{out}^\dagger[v])^n \ket{\text{vac}} \\
        =  \hat{B}(\phi) \hat{R}_\text{out}(\pi)\ket{\Psi_{\text{out}}},
    \end{aligned}
\end{equation}
where 
\begin{equation}
    \ket{\Psi_{\text{out}}} = \sum_n c_n  (\hat{a}_\text{out}^\dagger[v])^n \ket{\text{vac}} = e^{-i\hat{H}_B T}\ket{\Psi_\text{in}}
\end{equation}
represents the ligh state after free propagation for time $T$, which is given by replacing $\hat{a}_\text{in}[v]$ in $\ket{\Psi_\text{in}}$ with $\hat{a}_\text{out}[v]$.
This shows that the empty cavity gives a $\pi$-phase shift, followed by the optical loss represented by $\hat{B}(\phi)$.

\subsection{Reflection-based conditional displacement gate}
We now identify the unitary dynamics of RCD gates, which is given by $\hat{H}(t)$~\eqref{ap_eq:total_eff_H}.
To remove the time-dependent term of the Hamiltonian in advance, we apply a time-dependent unitary transformation with
\begin{equation} \label{ap_eq:U_d}
    \begin{aligned}
        \hat{U}_\text{d}(t) =& \hat{U}_{c,\text{d}}(t) \Pi_{j=0,1} \hat{U}_{j,\text{d}}(t), \\
        \hat{U}_{c,\text{d}}(t) =& e^{\hat{\sigma}_x [b(t) \hat{c}^\dagger - \text{h.c.}]}, \\
        \hat{U}_{j,\text{d}}(t) =& \exp\ab\{\hat{\sigma}_x \int \dd{\omega} \zeta_j(\omega,t) \hat{a}_j^\dagger(\omega) - \text{h.c.} \},
    \end{aligned}
\end{equation}
where $b(t)$ is a free paramter fulfilling $b(0) = 0$, and 
\begin{equation}
    \zeta_j(\omega,t) =  \sqrt{\frac{\kappa_j}{\pi}} \int_0^t \dd{t^\prime} b(t^\prime) e^{-i\omega(t-t^\prime)}.
\end{equation}
This gives the transformed Hamiltonian (see Ref.~\cite{Kikura2025} for the detailed derivation)
\begin{equation}
    \begin{aligned}
        \hat{H}_\text{d}(t) =& \hat{U}_\text{d}^{-1}(t) \hat{H}(t) \hat{U}_\text{d}(t) + i\ab[\odv{\hat{U}_\text{d}^{-1}(t)}{t}]\hat{U}_\text{d}(t) \\
        =& \hat{H}(t) - \hat{\sigma}_x \ab\{i[\dot{b}(t) + \kappa b(t)] \hat{c}^\dagger + \text{h.c.}\}.
    \end{aligned}
\end{equation}
Thus, setting the free parameter $b(t)$ such that 
\begin{equation}
    i[\dot{b}(t) + \kappa b(t)] = \lambda(t)
\end{equation}
gives
\begin{equation}
    \hat{H}_\text{d} = \hat{H}_B + \hat{H}_\text{int},
\end{equation}
which is the time-independent Hamiltonian for an empty cavity.
This shows that the unitary propagator of $\hat{H}(t)$ is given by
\begin{equation}
    \hat{U}(t) = \hat{U}_\text{d}(t) e^{-i\hat{H}_\text{d} t}.
\end{equation}
At a gate-terminating time $T$ such that $b(T) = \lambda(T) = 0$, we find
\begin{equation}
    \int \dd{\omega} \zeta_j(\omega, T) \hat{a}_j^\dagger(\omega) = 
    \begin{cases}
        \sqrt{2\kappa_\text{ex}} \int_0^T \dd{t}b(t) \hat{a}_\text{out}^{\text{(s)}}(t) & \text{if $j=0$}, \\
        \sqrt{2\kappa_\text{in}} \int_0^T \dd{t}b(t) \hat{a}_\text{loss}^{\text{(s)}}(t) & \text{if $j=1$},
    \end{cases}
\end{equation}
and thus
\begin{equation}
    \hat{U}(T) = e^{\hat{\sigma}_x \int \dd{t} \ab\{b(t) [\sqrt{2\kappa_\text{ex}}\hat{a}_\text{out}^{\text{(s)}}(t) + \sqrt{2\kappa_\text{in}}\hat{a}_\text{loss}^{\text{(s)}}(t)] - \text{h.c.} \} } e^{-i\hat{H}_\text{d} T}.
\end{equation}

To perform a conditional displacement gate of amplitude $\alpha$ to the wavepacket mode $v$, we set
\begin{equation}
    b(t) = \frac{\alpha v(t)}{\sqrt{2\kappa_\text{ex}}},
\end{equation}
leading to
\begin{equation}
    \lambda(t) = \frac{i\alpha}{\sqrt{2\kappa_\text{ex}}} [\dot{v}(t)+\kappa v(t)].
\end{equation}
This results in
\begin{equation}
    \hat{U}(T) = e^{\hat{\sigma}_x [\alpha \hat{a}_\text{out}^\dagger[v]-\text{h.c.}]} e^{\hat{\sigma}_x [\sqrt{\eta_\text{ex}^{-1}-1}\alpha \hat{a}_\text{loss}^\dagger[v]-\text{h.c.}]} e^{-i\hat{H}_\text{d} T}.
\end{equation}
Therefore, for a sufficiently long pulse $v(t)$, the gate action is represented by
\begin{equation}
    \hat{U}(T) = \text{CD}_\text{out}(\alpha)\text{CD}_\text{loss}(\sqrt{\eta_\text{ex}^{-1}-1}\alpha) \hat{B}(\phi) \hat{R}_\text{out}(\pi),
\end{equation}
where conditional displacement operators are defined by
\begin{equation}
    \text{CD}_\mu(\alpha) = e^{\hat{\sigma}_x (\alpha \hat{a}_\mu^\dagger[v]-\alpha^\ast \hat{a}_\mu[v])} \quad (\mu \in \{\text{``out'', ``loss''}\}).
\end{equation}
In the following, we cancel the $\pi$ phase shift $\hat{R}_\text{out}(\pi)$ by offsetting the phase origin by $\pi$ after the reflection.

Considering that the loss-mode state---which is initially a vacuum state---is not accessible, the output state is given by
\begin{equation}
    \begin{aligned}
        \hat{\rho}(T) =& \Tr_\text{loss}\ab[\hat{U}_\text{d}(T)\hat{B}(\phi) (\hat{\rho}_{\text{out}} \otimes \ketbra{0}[_\text{loss}]{0})\hat{B}^\dagger(\phi)\hat{U}_\text{d}^\dagger(T)] \\
        =& \text{CD}_\text{out}(\alpha) \mathcal{E}(\hat{\rho}_{\text{out}})\text{CD}_\text{out}^\dagger(\alpha),
    \end{aligned}
\end{equation}
with
\begin{equation}
    \begin{gathered}
        \mathcal{E}(\hat{\rho}) = \Tr_\text{loss}[\hat{U}_\text{loss}(\alpha, \eta_\text{ex}) (\hat{\rho} \otimes \ketbra{0}[_\text{loss}]{0}) \hat{U}_\text{loss}^\dagger(\alpha, \eta_\text{ex})], \\
        \hat{U}_\text{loss}(\alpha, \eta_\text{ex}) = \text{CD}_\text{loss}(\sqrt{\eta_\text{ex}^{-1}-1}\alpha)\hat{B}(\phi).
    \end{gathered}
\end{equation} 
For the ideal case, $\kappa_\text{in}=0$, we find $\hat{U}_\text{loss}(\alpha,\eta_\text{ex}) = \hat{I}$, resulting in the ideal RCD gate.
In the main text, we may follow standard convention and omit the subscripts ``in'' and ``out'', for notational simplicity.

\supplsection{Atomic-decay effect} \label{ap:atomic-decay_effect}

Here, we estimate the probability of atomic spontaneous decay at rate $\gamma$ through RCD-gate dynamics.
For the atom in ground states, the atomic decay is induced by two processes~\cite{Kikura2025}: (i) The atom is excited by the classical laser and subsequently decays. The instantaneous effective decay rate is approximately $\gamma |\Omega(t)/\Delta|^2$. (ii) The atom is excited by the cavity field and subsequently decays. The instantaneous effective decay rate is approximately $\gamma |g/\Delta|^2 \aab*{\hat{c}^\dagger(t)\hat{c}(t)}$.
While the latter decay is suppressed arbitrarily by increasing the detuning $\Delta$, the former does not depend on $\Delta$ since $|\Omega(t)/\Delta| = |\lambda(t)/g|$.
Thus, for a sufficiently large $\Delta$, the effective decay rate is given by~\cite{Kikura2025}
\begin{equation}
    \gamma_\text{eff}(t) \simeq \gamma \vab{\frac{\lambda(t)}{g}}^2.
\end{equation}
The atomic-decay probability is then given by
\begin{equation}
    p_\text{sp} \simeq 1-e^{-\int \dd{t} 2\gamma_\text{eff}(t)}.
\end{equation} 
From the relation
\begin{equation}
    |\lambda(t)|^2 = \frac{|\alpha|^2}{2\kappa_\text{ex}} \ab[\vab{\dot{v}(t)}^2 + \kappa \odv*{\vab{v(t)}^2}{t} + \kappa^2 \vab{v(t)}^2],
\end{equation}
we find
\begin{equation}
    \int \dd{t} \vab{\lambda(t)}^2 = \frac{\kappa^2 \vab{\alpha}^2}{2\kappa_\text{ex}} \ab[1 + \frac{\int \dd{t} \vab{\dot{v}(t)}^2}{\kappa^2}],
\end{equation}
resulting in
\begin{equation}
    p_\text{sp} \simeq 1- \exp\ab\{-\frac{|\alpha|^2}{2\eta_\text{ex}C}\ab[1 + \frac{\int \dd{t} \vab{\dot{v}(t)}^2}{\kappa^2}] \}.
\end{equation}

\supplsection{Finite-length-pulse effect in reflection off an empty cavity} \label{ap_sec:finite-pulse_effect}

In Sec.~\ref{ap_sec:RCD_gates}, we show that the reflection off an empty cavity is denoted by a beamsplitter operator $\hat{B}(\phi)$ for a sufficiently long pulse $v(t)$.
In general, however, the reflected pulse shape is disturbed from the input one, due to the frequency dependence of the cavity response~\eqref{ap_eq:r(omega)_l(omega)}.
Here, we quantitatively evaluate finite-length-induced errors.

For calculation simplicity, we consider the initial light state to be a coherent state with amplitude $\beta$, where the average photon number is $\bar{n}_\text{in} = \vab{\beta}^2$.
From Eq.~\eqref{ap_eq:Psi(T)_empty_cavity}, the reflected state is 
\begin{equation}
    \begin{aligned}
        &\exp\ab\{ \int\dd{\omega} [\beta r(\omega)v(\omega)e^{-i\omega T} \hat{a}_0^\dagger(\omega) -\text{h.c.}] \} \\
        &\times \exp\ab\{ \int\dd{\omega} [\beta l(\omega)v(\omega)e^{-i\omega T} \hat{a}_1^\dagger(\omega) -\text{h.c.}] \} \ket{\text{vac}}.
    \end{aligned}
\end{equation}
This leads to the fidelity with the long-pulse-limit state as
\begin{equation}
    \begin{aligned}
        |\bra{\text{vac}}& e^{\int\dd{\omega} \{\beta[r(\omega)-r(0)]v(\omega)e^{-i\omega T} \hat{a}_0^\dagger(\omega) -\text{h.c.}\} } \\
        &\times e^{\int\dd{\omega} \{\beta[l(\omega)-l(0)]v(\omega)e^{-i\omega T} \hat{a}_1^\dagger(\omega) -\text{h.c.}\} } \ket{\text{vac}}|^2.
    \end{aligned}
\end{equation}
By using the commutation relation $[\hat{a}_j(\omega),\hat{a}_j^\dagger(\omega^\prime)] = \delta(\omega-\omega^\prime)$ and the Baker-Campbell-Hausdorff formula, the fidelity is reduced to
\begin{equation}
    \begin{aligned}
        &\exp\ab\{- \bar{n}_\text{in}\int \dd{\omega} \ab[\vab{r(\omega)-r(0)}^2 +\vab{l(\omega)-l(0)}^2] \vab{v(\omega)}^2  \ab\} \\
        &= \exp\ab\{-\bar{n}_\text{in} \int \dd{\omega} 4\eta_\text{ex}\ab[1- \frac{1}{(\omega/\kappa)^2 + 1}] \vab{v(\omega)}^2 \} \\
        &= \exp\ab\{-4\eta_\text{ex} \bar{n}_\text{in}  \ab[1- \int \dd{\omega}\frac{\vab{v(\omega)}^2}{(\omega/\kappa)^2 + 1}] \}, 
    \end{aligned}
\end{equation}
since $\int \dd{\omega} \vab{v(\omega)}^2 = 1$.
In the following, we consider a Gaussian pulse,
\begin{equation}
    v(t) = \frac{1}{(\pi \tau^2)^{1/4}}e^{-t^2/(2\tau^2)}, \quad v(\omega) = \frac{1}{(\pi \mathcal{B}^2)^{1/4}}e^{-t^2/(2\mathcal{B}^2)},
\end{equation}
where $\tau$ characterizes the pulse length and $\mathcal{B} = 1/\tau$ characterizes its bandwidth.
In this case, we find~\cite{Ng1969}
\begin{equation}
    \int_{-\infty}^{\infty}\dd{\omega}\frac{\vab{v(\omega)}^2}{(\omega/\kappa)^2 + 1} = \sqrt{\pi} \kappa \tau e^{(\kappa \tau)^2} \erfc(\kappa\tau),
\end{equation}
where $\erfc(x) = (2/\sqrt{\pi})\int_x^{\infty} e^{-t^2}\dd{t}$ is the complementary error function.
Thus, the infidelity induced by the finite-pulse effect is given by
\begin{equation}
    \begin{aligned}
        \epsilon_{\text{pulse}} =& 1 - e^{-4\eta_\text{ex} \bar{n}_\text{in} [1-\sqrt{\pi} \kappa \tau e^{(\kappa \tau)^2} \erfc(\kappa\tau)]} \\
        =& \frac{2\eta_\text{ex} \bar{n}_\text{in}}{(\kappa \tau)^2} + \mathcal{O}\ab(\frac{1}{(\kappa \tau)^4}).
    \end{aligned}
\end{equation}

\supplsection{Numerical simulation of RCD gates} \label{ap:numerical_simulation_method}

To simulate the interaction between an FLS inside a cavity and an itinerant quantum state in a wavepacket mode, we use an efficient method proposed by Kiilerich and M{\o}lmer~\cite{Kiilerich2019}.
In this model, a virtual cavity with the time-dependent complex coupling
\begin{equation}
    g_\text{in}(t) = \frac{v^\ast(t)}{\sqrt{1-\int_0^t \dd{t^\prime} |v(t^\prime)|^2}}
\end{equation}
releases an initial quantum state inside the cavity into the transmission line, as the state of the wavepacket mode $v$.
After the light reflects off the local system, the light is captured by another virtual cavity with coupling
\begin{equation}
    g_\text{out}(t) = -\frac{v^\ast(t)}{\sqrt{\int_0^t \dd{t^\prime} |v(t^\prime)|^2}}.
\end{equation}
This model gives the master equation for the quantum state $\hat{\rho}_\text{iso}$ consisting of the local system and the input and output virtual cavities as follows:
\begin{equation} \label{eq:master_eq_for_virtual_cavity}
    \odv{\hat{\rho}_\text{iso}}{t} = -i[\hat{H}_\text{iso},\hat{\rho}_\text{iso}] + \mathcal{D}[\hat{L}_\text{iso}]\hat{\rho}_\text{iso} + \sum_{j}\mathcal{D}[\hat{L}_j]\hat{\rho}_\text{iso},
\end{equation}
with
\begin{equation}
    \begin{aligned}
        \hat{H}_\text{iso}(t) =& \hat{H}_\text{sys}(t) + \frac{i}{2}\Big[\sqrt{2\kappa_\text{ex}}g_\text{in}(t)\hat{a}_\text{in}^\dagger\hat{c} \\
        & +\sqrt{2\kappa_\text{ex}}g^\ast_\text{out}(t)\hat{c}^\dagger\hat{a}_\text{out}+g_\text{in}(t)g_\text{out}^\ast(t)\hat{a}_\text{in}^\dagger \hat{a}_\text{out}- \text{h.c.} \Big],
    \end{aligned}
\end{equation}
where $\hat{a}_\text{in(out)}$ represents the annihilation operator of the input (output) virtual cavity mode.
Here, the damping terms $\mathcal{D}[\hat{L}]\hat{\rho} \coloneqq \hat{L} \hat{\rho} \hat{L}^\dagger - \{\hat{L}^\dagger \hat{L},\hat{\rho} \}/2$ include the Lindblad operator
\begin{equation}
    \hat{L}_\text{iso}(t) = \sqrt{2\kappa_\text{ex}}\hat{c} + g_\text{in}^\ast(t) \hat{a}_\text{in} + g_\text{out}^\ast(t) \hat{a}_\text{out},
\end{equation}
representing the loss from the output cavity due to the mode mismatch, and $\hat{L}_j$ representing the decay of the system, such as cavity internal loss and atomic decay.

\subsection{Coherent-state input}
Here, we consider that the input light is a coherent state.
This allows us to translate the input-light effect into the classical driving of the system---known as the Mollow transformation~\cite{Mollow1975}---and erase the input virtual cavity, reducing the dimension of the total Hilbert space~\cite{Kiilerich2020}.
In the following, we derive a reduced master equation for the local system and the output cavity.

Considering the damping rate of the input cavity at $|g_\text{in}(t)|^2/2$, the state inside the cavity remains a coherent state with a damped amplitude, and the total quantum state is given as $\hat{\rho}_\text{iso} = \ketbra{\beta(t)}[_\text{in}]{\beta(t)} \otimes \hat{\rho}_\text{so}$ with $\beta(t) = \beta \exp\ab(-\int_0^t \dd{t^\prime}|g_\text{in}(t^\prime)|^2/2)$.
To derive the differential equation for $\hat{\rho}_\text{so} = \Tr_\text{in}[\hat{\rho}_\text{iso}]$, we trace out the input cavity mode in Eq.~\eqref{eq:master_eq_for_virtual_cavity}.
By using $\hat{a}_\text{in}\ket{\beta(t)}_\text{in} = \beta(t)\ket{\beta(t)}_\text{in}$ and $g_\text{in}^\ast(t)\beta(t) = \beta v(t)$, we find
\begin{equation}
    \Tr_\text{in}\ab[[\hat{H}_\text{iso},\hat{\rho}_\text{iso}]] = [_\text{in}\!\braket*[3]{\beta(t)}{\hat{H}_\text{iso}}{\beta(t)}_\text{in},\hat{\rho}_\text{so}],
\end{equation}
with
\begin{equation}
    \begin{aligned}
        _\text{in}\!\braket*[3]{\beta(t)}{\hat{H}_\text{iso}}{\beta(t)}_\text{in} =& \hat{H}_\text{sys}(t) + \frac{i}{2}\Big\{ \sqrt{2\kappa_\text{ex}}g_\text{out}^\ast(t)\hat{c}^\dagger\hat{a}_\text{out} \\
        &+ [\beta v(t)]^\ast\hat{L}_\text{so}(t) -\text{h.c.} \Big\},
    \end{aligned}
\end{equation}
and 
\begin{equation}
    \begin{aligned}
        \Tr_\text{in}\ab[\mathcal{D}[\hat{L}_\text{iso}]\hat{\rho}_\text{iso}]=& \mathcal{D}[\hat{L}_\text{so}]\hat{\rho}_\text{so}  \\
        &+ \frac{1}{2}\ab[[\beta v(t)]^\ast \hat{L}_\text{so}(t) - \beta v(t) \hat{L}_\text{so}^\dagger(t), \hat{\rho}_\text{so}],
    \end{aligned}
\end{equation}
with 
\begin{equation}
    \hat{L}_\text{so} = \sqrt{2\kappa_\text{ex}}\hat{c} + g_\text{out}^\ast(t) \hat{a}_\text{out}.
\end{equation}
Thus, we obtain the reduced master equation
\begin{equation} \label{ap_eq:master_eq_for_rho_so}
    \odv{\hat{\rho}_\text{so}}{t} = -i[\hat{H}_\text{so},\hat{\rho}_\text{so}] + \mathcal{D}[\hat{L}_\text{so}]\hat{\rho}_\text{so}  + \sum_{j}\mathcal{D}[\hat{L}_j]\hat{\rho}_\text{so},
\end{equation}
where the reduced Hamiltonian is given by
\begin{equation}
    \begin{aligned}
        \hat{H}_\text{so}(t) =& \hat{H}_\text{sys}(t) + \frac{i}{2}\Big\{\sqrt{2\kappa_\text{ex}}g_\text{out}^\ast(t)\hat{c}^\dagger\hat{a}_\text{out} \\
        &+ 2[\beta v(t)]^\ast\hat{L}_\text{so}(t) -\text{h.c.} \Big\}.
    \end{aligned}
\end{equation}

The authors in Ref.~\cite{Kiilerich2020} derived a reduced master equation without an output virtual cavity, which is given by removing the terms for the output virtual cavity in Eq.~\eqref{ap_eq:master_eq_for_rho_so},
\begin{equation} \label{ap_eq:master_eq_for_rho_s}
    \odv{\hat{\rho}_\text{s}}{t} = -i[\hat{H}_\text{s},\hat{\rho}_\text{s}] + \mathcal{D}[\hat{L}_\text{s}]\hat{\rho}_\text{s}  + \sum_{j}\mathcal{D}[\hat{L}_j]\hat{\rho}_\text{s},
\end{equation}
with
\begin{gather}
    \hat{H}_\text{s}(t) = \hat{H}_\text{sys}(t) + i\{[\beta v(t)]^\ast\hat{L}_s - \beta v(t) \hat{L}_s^\dagger \}, \\
    \hat{L}_\text{s} = \sqrt{2\kappa_\text{ex}} \hat{c}.
\end{gather}
The solution of the reduced master equation~\eqref{ap_eq:master_eq_for_rho_s} gives the information of the scattered field from the local system.
For example, the intensity of the scattered field is given as
\begin{equation} \label{ap_ex:I_out}
    I_\text{out}(t) = \Tr\ab[[\hat{L}_s + \beta v(t)]^\dagger [\hat{L}_s + \beta v(t)] \hat{\rho}_s(t)].
\end{equation}

For full-level numerical simulation, we employ the system Hamiltonian presented in Eq.~\eqref{ap_eq:full-level_H_sys} and Lindblad operators that describe cavity loss and atomic decay,
\begin{equation}
    \begin{aligned}
        \hat{L}_1 =& \sqrt{2\kappa_\text{in}}\hat{c}, \\
        \hat{L}_2 =& \sqrt{2r_1\gamma} \ketbra{0}[_\text{q}]{e_1}, \quad \hat{L}_2 = \sqrt{2(1-r_1)\gamma} \ketbra{1}[_\text{q}]{e_2}, \\
        \hat{L}_3 =& \sqrt{2r_2\gamma} \ketbra{1}[_\text{q}]{e_2}, \quad \hat{L}_2 = \sqrt{2(1-r_2)\gamma} \ketbra{0}[_\text{q}]{e_2},
    \end{aligned}
\end{equation}
where $r_j\,(j\in \{1,2\})$ represents the branching ratio, which is set as 0.5 in our numerical simulations.
For the results shown in Fig.~\ref{fig2}(b), we numerically solve the master equation~\eqref{ap_eq:master_eq_for_rho_s} and calculate the intensity $I_\text{out}(t)$~\eqref{ap_ex:I_out}, by using QuTip 5~\cite{QuTip5}.

For the effective-model simulation presented in Figs.~\ref{fig2} and \ref{fig3}, we use the effective Hamiltonian~\eqref{ap_eq:H_sys^eff} and further replace the Lindblad operators of atomic decay with the effective ones~\cite{Kikura2025},
\begin{equation}
    \begin{aligned}
        \hat{L}_2^\text{eff} =& -\frac{\sqrt{2r_1\gamma}}{g}[\lambda(t) \ketbra{0}[_\text{q}]{1} -\chi \ketbra{0}[_\text{q}]{0}\hat{c}]  ,\\
        \hat{L}_3^\text{eff} =& -\frac{\sqrt{2(1-r_1)\gamma}}{g} [\lambda(t)\ketbra{1}{1} - \chi \ketbra{1}[_\text{q}]{0}\hat{c}],\\
        \hat{L}_4^\text{eff} =& -\frac{\sqrt{2r_2\gamma}}{g}[\lambda(t) \ketbra{1}[_\text{q}]{0} -\chi \ketbra{1}[_\text{q}]{1}\hat{c}] ,\\
        \hat{L}_5^\text{eff} =& -\frac{\sqrt{2(1-r_2)\gamma}}{g} [\lambda(t) \ketbra{0}[_\text{q}]{0} -\chi \ketbra{0}[_\text{q}]{1}\hat{c}],
    \end{aligned}
\end{equation}
where we have used an effective operator formalism~\cite{Reiter2012}.

\bibliography{refs}
\end{document}